\documentclass[prd,preprint,tightenlines,floatfix,showpacs,preprintnumbers,nofootinbib,draft]{revtex4}
 \usepackage[dvips,final]{graphicx}
  \usepackage{amssymb}
   \usepackage{amsmath}
    \usepackage{epsfig}
     \usepackage{bm}
      \usepackage{pifont}
       \usepackage[dvips,final]{graphicx}

\begin{document}
\thispagestyle{empty}
\preprint{RUB-TPII-09/04}

\title{Deep inside the pion. Reconciling QCD
       theory with data}

\author{A.~P.~Bakulev}
 \email{bakulev@thsun1.jinr.ru}
  \affiliation{Bogoliubov Laboratory of Theoretical Physics, JINR,
   141980 Dubna, Russia}

\author{S.~V.~Mikhailov}
 \email{mikhs@thsun1.jinr.ru}
  \affiliation{Bogoliubov Laboratory of Theoretical Physics, JINR,
   141980 Dubna, Russia}

\author{N.~G.~Stefanis}
 \email{stefanis@tp2.ruhr-uni-bochum.de}
  \affiliation{Institut f\"{u}r Theoretische Physik II,\\
  \vspace{10mm}
   Ruhr-Universit\"{a}t Bochum,
   D-44780 Bochum, Germany\\ \vspace{10mm}
      \textsl{\em Dedicated to Prof. Klaus Goeke on the occasion
                  of his 60th birthday.}}

\begin{abstract}
Recent developments in the QCD description of the pion structure
are reviewed. The CLEO pion-photon transition data analysis favors
a distribution amplitude for the pion that is double-humped but
endpoint-suppressed. After a short outline of the derivation of
this amplitude from QCD sum rules with nonlocal condensates, we
present the fully fledged analysis of the CLEO data prefaced by
predictions for the $F^{\gamma\rho\pi}$ form factor and commenting
on the inherent theoretical uncertainties due to higher twists and
NNLO perturbative corrections. We supplement our discussion by
considering within QCD factorization theory, the electromagnetic
pion form factor at NLO accuracy on one hand, and diffractive
di-jets production on the other, comparing our predictions with
the respective experimental data from JLab and the Fermilab E791
collaboration. In all cases, the agreement is impressive.
\end{abstract}

\maketitle
\section{Introduction}
Advances in theoretical science are mostly based on finding useful
and compact descriptions of a phenomenon of interest.
Understanding the nonperturbative dynamics of quarks and gluons
inside the pion is a question too difficult to be addressed from
first principles within QCD. In order to gain some insight into
the nonperturbative physics of the QCD vacuum, the concept of
nonlocal condensates was introduced \cite{MR86,BR91,MS93}.

In this framework, a simple and compact descriptor of the complex
vacuum structure is provided by the average virtuality $\langle
k_{q}^{2} \rangle = \lambda_q^2$ of vacuum quarks, whose inverse
$\lambda_{q}^{-1}$ defines the correlation length of the scalar
quark nonlocal condensate. The nonlocality parameter has been
estimated using QCD sum rules \cite{BI82} and lattice calculations
\cite{DDM99}; just recently \cite{BMS02} it has been extracted
from the data of the CLEO collaboration \cite{CLEO98} on the
pion-photon transition. One finds values in the range
$\lambda_{q}^{2} =
 \langle \bar{q} \left(i g\,\sigma_{\mu \nu}G^{\mu \nu} \right) q
 \rangle / (2\langle \bar{q} q \rangle)=(0.35-0.5)~{\rm GeV}^2
$ which pertain to a correlation length varying, respectively,
between $0.33$~fm and $0.28$~fm, while the CLEO data favors the
value $\lambda_{q}^{2} = 0.4$~GeV$^2$ \cite{BMS02} and a
correlation length of about $0.31$~fm.

Using QCD sum rules within this framework and a Gaussian ansatz
with the single parameter $\lambda_q^2$ for the vacuum distributions,
the pion distribution amplitude (DA) was first
derived in \cite{MR86,BM98}. Later, this approach was refurbished
and rectified \cite{BMS01}, providing constraints on the first ten
moments of the pion DA and an estimate for the inverse moment
using an \emph{independent} sum rule. With recourse to the fast
decrease of the moment values with its number $N$,
\cite{BMS01mor,BMS03dur}, the Gegenbauer coefficients of the pion
DA were calculated within uncertainty ranges and it was found that
only the first two of them $a_2$ and $a_4$ are important; the rest
are negligible \cite{BMS01,BMS03dur}. This gives us a handy tool
to reconstruct the pion distribution amplitude \cite{BMS01} and
calculate with it pion observables, like the pion-photon
transition form factor $F_{\pi\gamma^*\gamma^*}(Q^2)$
\cite{BMS02,BMS03,BMS03efr} and the electromagnetic form factor
$F_{\pi}(Q^2)$ \cite{BPSS04}. Confronting this pion DA (in terms
of the first two Gegenbauer coefficients) with the constraints
extracted from the CLEO data \cite{CLEO98} via a best-fit analysis
\cite{SY99}, it was found \cite{BMS02,BMS03} that it is within the
$1\sigma$-error ellipse while the exclusion of the
Chernyak-Zhitnitsky (CZ) \cite{CZ84} distribution amplitude at the
$4\sigma$ level and of the asymptotic one at the $3\sigma$ level
was reinforced.

\section{Nonlocal condensates and the pion distribution amplitude}
\begin{figure}[b]
\centerline{\includegraphics[width=\textwidth]{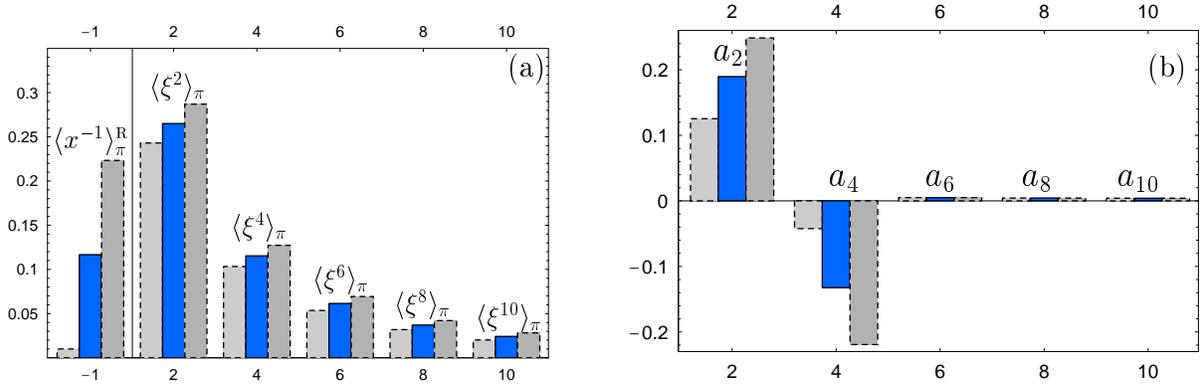}}%
  \vspace*{-2mm}\caption{ \footnotesize
  (a) First ten ($N=10$) nonzero moments,
   determined with nonlocal QCD sum rules \cite{BMS01},
   $\langle{\xi^N}\rangle_\pi$  and
   $\langle x^{-1}\rangle^{\rm R}_{\pi} =(1/3)
   \langle x^{-1}\rangle_{\pi} - 1$
   (the superscript R meaning ``reduced'')
   of $\varphi_{\rm BMS}$ (dark blue bars) together with their upper
   and lower error-bars as light-grey bars.
   (b) Histogram of the first nonzero Gegenbauer coefficients $a_n$
   of the BMS pion DA and the envelopes of the ``bunch'' as light-grey
   bars.}
 \label{fig:histogram}
\end{figure}
The pion DA is a universal process-independent characteristic of the
pion, explaining how the longitudinal momentum $P$ is partitioned
between its two valence partons quark ($x$) and antiquark
($\bar{x}=1-x$), this being a reflection of the underlying
nonperturbative dynamics. To leading twist-2, one has
\begin{equation}
  {\langle 0\mid \bar{d}(z) \gamma^{\mu}\gamma_{5} {\cal C}(z,0) u(0)
            \mid \pi(P)
  \rangle}\Big|_{z^2=0}
=
  i f_{\pi} P^{\mu} \int^{1}_{0} dx {\rm e}^{ix(zP)}
  \varphi_{\pi}\left(x,\mu_{0}^{2} \right) \, ,
\label{eq:def-phi-pi}
\end{equation}
where $ {\cal C}(0,z)
  = {\cal P}
  \exp\!\left[-ig_s\!\!\int_0^z t^{a} A_\mu^{a}(y)\, dy^\mu\right]
$ is the path-ordered phase factor (the connector \cite{Ste84}) to
preserve gauge invariance, $f_{\pi}$ is the pion decay coupling,
and the normalization is $\int^{1}_{0} dx
\varphi_{\pi}\left(x,\mu^{2}\right)=1$.

To get a handle on the pion DA, the aim is to relate it with the
nonperturbative QCD vacuum in terms of nonlocal condensates. This
is achieved by relating the pion and its first resonance by means
of a sum rule, based on the correlator of two axial currents:
\begin{eqnarray}
  f_{\pi}^2\varphi_\pi(x) +
  f_{A_1}^2\varphi_{A_1}(x)\exp\left(-\frac{m^2_{A_1}}{M^2}\right)
&=&
   \int_{0}^{s_{\pi}^0}\rho^{\rm pert}_{\rm NLO}(x;s)e^{-s/M^2}ds
  +
   \frac{\langle \alpha_{\rm s} GG \rangle}{24\pi M^2}\
   \Phi_{\rm G}\left(x;M^2\right) \nonumber \\
&+&  \frac{8\pi\alpha_{\rm s}\langle{\bar{q}q\rangle}^2}{81M^4}
      \sum_{i={\rm S,V,T_{1,2,3}}}\Phi_i\left(x;M^2\right) \, .
\label{eq:nlcsrda}
\end{eqnarray}
Here the index $i$ runs over scalar (S), vector (V), and tensor
(T) condensates, $M^2$ is the Borel parameter, and $s_{\pi}^0$ is
the duality interval in the axial channel, whereas $\rho^{\rm
pert}_{\rm NLO}(x;s)$ is the spectral density in NLO perturbation
theory \cite{BMS01,BM98}.
 Above, the dependence on the
crucial non-locality parameter $\lambda_q^2$ enters the sum rule
in the way exemplified by the numerically important scalar-condensate
contribution
\begin{eqnarray}
\label{PhiS}
 \Phi_S\left(x;M^2\right)
  &=& \frac{18}{\bar\Delta\Delta^2}
       \Bigl\{
        \theta\left(\bar x>\Delta>x\right)
         \bar x\left[x+(\Delta-x)\ln\left(\bar x\right)\right]
       +  \left(\bar x\rightarrow x\right) + \nonumber \\
&&\qquad\quad
       + \theta(1>\Delta)\theta\left(\Delta>x>\bar\Delta\right)
         \left[\bar\Delta
              +\left(\Delta-2\bar xx\right)\ln(\Delta)\right]
         \Bigr\}
\end{eqnarray}
with $\Delta \equiv\lambda_q^2/(2M^2)$, $\bar\Delta\equiv
1-\Delta$.
One appreciates from this last
expression that neglecting the vacuum correlation length (as in
the approach of \cite{CZ84}), the end-point contributions ($x\to
0$ or $1$) are strongly enhanced by $\delta(x), \delta'(x) \ldots$
because for $\lambda_q^2 \to 0$ one obtains $
  \lim\limits_{\Delta \to 0}\Phi_S\left(x;M^2\right)
=
  9\left[ \delta(x)+  \delta(1-x)\right].
$
By virtue of the finiteness of $\lambda_q^2$, the sum rule
(\ref{eq:nlcsrda}) can supply us with constraints on the first ten
moments $\langle\xi^N\rangle_\pi \equiv \int_{0}^{1}
\varphi_\pi(x)(2x-1)^N dx$ of the pion DA that are decreasing with
increasing polynomial order to zero, i.e.,
$\langle \xi^N\rangle\to [3/(N+1)(N+3)]$
(see Fig.\ \ref{fig:histogram}
(a)). In addition, we can obtain an \emph{independent} sum rule to
constraint also the inverse moment $\langle x^{-1}\rangle_{\pi}
\equiv \int_{0}^{1}\varphi_\pi(x)\ x^{-1} dx $ quite accurately
\cite{BMS01,BMS01mor} -- see Fig.\ \ref{fig:histogram} (a). This
is qualitatively how one derives correlated values of the first
two Gegenbauer coefficients $a_2$ and $a_4$ in the (nonlocal) QCD
sum-rules picture, given that the higher ones are practically zero
(see Fig.\ \ref{fig:histogram} (b)). The final step is then to
model the pion DA according to the expression
\begin{equation}
  \varphi^{\rm BMS}(x; \mu_0^2)
= \varphi^{\rm as}(x)
       \left[1 + a_2(\mu_0^2)\ C^{3/2}_2(2x-1)
               +   a_4(\mu_0^2)\ C^{3/2}_4(2x-1) \right] \, ,
\label{optG}
\end{equation}
providing the ``bunch'' of DAs shown in Fig.\ 2 (Left) together with
the optimum sample, termed BMS model \cite{BMS01}
($a_2(\mu_0^2)=0.2$, $a_4(\mu_0^2)=-0.14$), in comparison with the
CZ and asymptotic pion DAs. Note that all DAs mentioned are
normalized at the same scale $\mu_0^2\simeq 1$~GeV$^2$. In Table
\ref{tab:variousNLSRs}, we compile the main features of our SRs,
contrasting them with previous ones.
\begin{table}[h]
\caption{Determining the pion DA from QCD nonlocal sum rules,
explaining the entries on its theoretical and phenomenological
side, as well as its outcome. The differences between our present
approach \cite{BMS01} and previous ones is pointed out.}
\begin{ruledtabular}
\begin{tabular*}{\textwidth}{%
|p{7.5mm}|cp{25mm}c|cp{30.5mm}c|cp{38mm}c|cp{45mm}c|}
{\hfil Ref.\hfil}&
     & {\hfil SR theor. part\hfil}&&
        & {\hfil SR phenom. part\hfil}&&
           & {\hfil SR estimates\hfil}&&
              & {\hfil Outcome\hfil}&\\ \hline \hline
{\hfil\protect{\cite{MR86}}$\vphantom{^{\int_0^1}_{\int_0^1}}$\hfil}&
     & {$O(\alpha_s)$ pQCD; quark, quark-gluon and gluon NLCs}&&
        & $\pi$-state $+$ conti\-nuum at $s_0\approx0.7~\text{GeV}^2$&&
           & {$\langle \xi^2 \rangle_{\pi}$,
           $\langle \xi^4 \rangle_{\pi}$,
           $\langle \xi^6 \rangle_{\pi}$}&&
              & {Pion DA should be closer to asymptotic rather than to
              CZ DA$\vphantom{_{\int_0^1}}$}&
                 \\ \hline
{\hfil\protect{\cite{BM98}}$\vphantom{^{\int_0^1}_{\int_0^1}}$\hfil}&
     & {\protect{\cite{MR86}} $+$ modified gluon NLC}&&
        & {$\pi$- and $A_1$-states $+$ continuum at
           $s_0 \approx 2.2$~GeV$^2$}&&
           & {$\langle \xi^2 \rangle_{\pi,A_1}, \ldots,
           \langle \xi^{10} \rangle_{\pi,A_1}$,
              $\langle 1/x \rangle_{\pi,A_1}$ with error\--bars}&&
              & {Two models of pion DAs: one includes the second
                 Gegenbauer harmonic, i.e., $a_2\neq 0$;
                 the other is end-point suppressed as
                 $(x\bar{x})^{2.3}\vphantom{_{\int_0^1}}$}&
                 \\ \hline
{\hfil\protect{\cite{BMS01}}$\vphantom{^{\int_0^1}_{\int_0^1}}$\hfil}&
    & {\protect{\cite{BM98}} $+$ corrected quark-gluon
        NLC\hspace{2mm}$T_1$}&&
    & {The same as in \protect{\cite{BM98}}: Borel window
       $[0.5,~2]$~GeV$^2$}&&
    & {$\langle \xi^2 \rangle_{\pi,A_1}, \ldots,
        \langle \xi^{10} \rangle_{\pi,A_1}$,
       $\langle 1/x \rangle_{\pi,A_1}$ with more conservative
         error-bars}&&
    & {``Bunch'' of self-consistent DAs (Fig.\ \ref{fig:GaRhoPi}(Left))
         with 2 Gegenbauer harmonics
        (Fig.\ \ref{fig:histogram}(b))$\vphantom{_{\int_0^1}}$}&
                  \\
\end{tabular*}
\end{ruledtabular}
\label{tab:variousNLSRs}
\end{table}

\section{Light-cone sum-rule predictions for $F^{\gamma^*\rho \pi}$ }
The $F^{\gamma\rho \pi}$ form factor appears as an inevitable part
of the $F^{\gamma \gamma^* \pi}$ transition form factor in a
light-cone sum-rule (LCSR) calculation. The main advantage of this
method is that one can calculate the form factor for sufficiently
large photon virtualities to obtain the perturbative spectral
density, and then analytically continue the result to the limit
$q^2 \simeq0$ using a dispersion relation. In this scheme,
$F^{\gamma\rho \pi}$ expresses the ``hadronic" content of the
quasi on-shell photon $\gamma(q^2)$ involved in the process
$\gamma^*(Q^2)~\gamma(q^2) \to \pi^0$. This calculational approach
has been proposed by Khodjamirian in \cite{Kho99} and
$F^{\gamma\rho \pi}(Q^2)$ was computed at the LO level of the LCSRs.

The form factor $Q^4 F^{\gamma\rho \pi}(Q^2)$ obtained in this
framework depends mainly on the \textit{differential} pion
characteristic
$\displaystyle \frac{d}{dx}\varphi_{\pi}(x)|_{x=\epsilon},~\epsilon
\sim \frac{s_{\rho}}{Q^2}$,
in an $\epsilon$-neighborhood of the origin.
This feature is opposite to the case of the
$Q^2 F^{\gamma \gamma^* \pi}(Q^2)$
form factor, which depends mainly on the inverse moment
$\langle x^{-1}\rangle_{\pi}=\int^1_0 \varphi_\pi(x;\mu^2){x}^{-1}dx$,
i.e., on an \textit{integral} pion characteristic \cite{BMS01,BMS03}.
From this point of view, $Q^4 F^{\gamma \rho \pi}(Q^2)$
can provide complementary information on the pion DA and help
discriminate among different pion DA models.
\begin{figure*}[t]
\centerline{\includegraphics[width=0.48\textwidth]{%
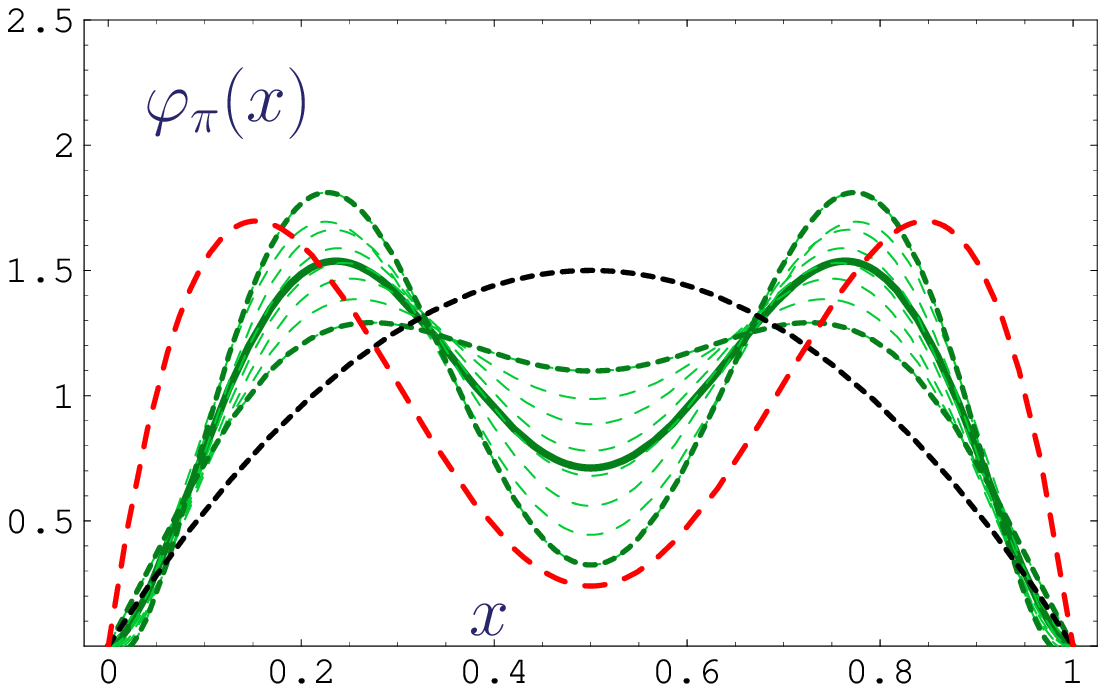}~%
  \includegraphics[width=0.48\textwidth]{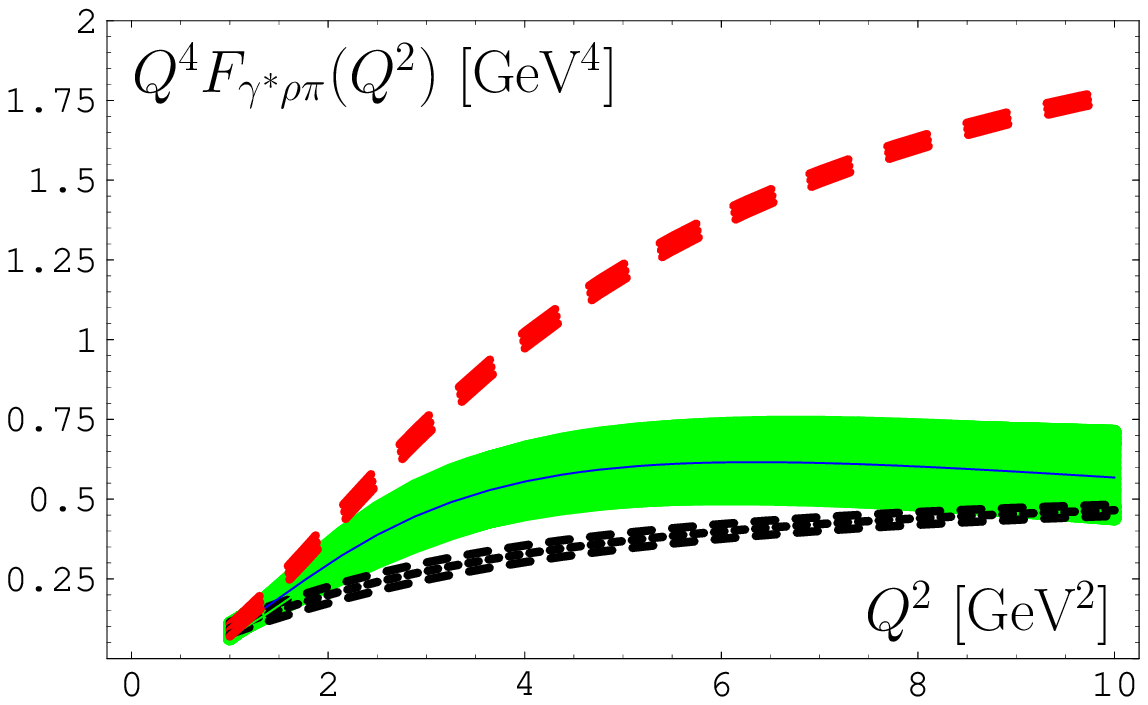}}
   \caption{\footnotesize \textbf{Left}:BMS ``bunch'' of the pion DAs
    contrasted with two extreme alternatives (asymptotic DA---dotted
    line and CZ model---long-dashed line) at
    $\mu^2 \approx 1$ GeV$^{2}$.
    \textbf{Right}: Predictions for $Q^4 F^{\gamma\rho \pi}(Q^2)$ for
    the pion DAs shown on the left. The thickness of the two broken
    lines corresponds to the variation of the twist-4 parameter
    in the range $\delta_{\rm Tw-4}^2=(0.15-0.23)$~GeV$^2$.
    \label{fig:GaRhoPi}}
\end{figure*}

Our predictions for $Q^4 F^{\gamma \rho \pi}(Q^2)$,
based on the BMS bunch, Fig.~2 (Left),
and on a complete NLO calculation for the corresponding spectral
density \cite{BMS02}, are presented in Fig.~2 (Right) in the form of a
shaded strip, with the central line denoting the BMS model.
These calculations are sketched in the next section.
Here, let us mention only the main features: (i) The
$\alpha_s$--corrections appear to be rather large, of the
order of $30\%$, and negative. (ii) The twist-4 contribution turns
out to be very important, larger than $30\%$ for $Q^2$ values
below 3~GeV$^2$ and also negative. (iii) An improved Breit-Wigner
ansatz for the phenomenological spectral density $\rho^{\rm mes}$ is
used that increases the result for the form factor by about $6\%$.

Comparing the different models in Fig.\ 2 (Left), one can
understand how the slope of $\varphi_{\pi}(x)$ in the domain $x
\sim s_{\rho}/(Q^2+s_{\rho})\approx 0.2 $ (at
$s_{\rho}=1.5$~GeV$^2, Q^2 \approx 6$ GeV$^{2}$) translates into
the curve for the corresponding $Q^4 F^{\gamma\rho \pi}(Q^2)$ form
factor in Fig.\ 2 (Right). All predictions shown are ``smeared''
curves, their thickness being a practical measure for the allowed
variation of the twist-4 parameter
$\delta_{\rm Tw-4}^2=(0.15-0.23)$~GeV$^2$.

\section{Analysis of the CLEO data}
Foremost among the many open questions in the nonperturbative
regime of QCD is the determination of the parameters to model the
shape of hadron DAs---prime examples being the pion Gegenbauer
coefficients, on focus here, and those for the nucleon DA (for a
review on the latter, see \cite{Ste99}). This task, though
obviously of paramount importance is noted for its intransigence.
However, the high-precision CLEO data \cite{CLEO98} on the
$\pi\gamma$ transition have improved the situation for the pion
significantly.

Indeed, Schmedding and Yakovlev (SY) \cite{SY99} have analyzed
this data set using a further extension to the NLO (of pQCD) of
the LCSR for the transition form factor
$F^{\gamma^*\gamma\pi}(Q^2,q^2 \approx 0)$, developed in
\cite{Kho99}. We adopted this approach in \cite{BMS02,BMS03}
improving it in the following respects: (i) A more accurate
point-to-point 2-loop ERBL \cite{ERBL79} evolution has been
employed,
      taking into account the quark thresholds.
(ii)  The contribution of the twist-4 term has been re-estimated to
      read
      $\delta^2(1{\rm GeV}^{2}) = (0.19\pm 0.02)~{\rm GeV}^{2}$ and the
      role of these uncertainties has been investigated in detail.
(iii) The procedure to determine the error range of the
      $1\sigma$- and $2\sigma$ error contours has been improved and
      uncertainties of high-order radiative corrections have been
      involved in the analysis.
For more detailed information and explicit expressions, the interested
reader may consult \cite{BMS02,BMS03}. In the present exposition we
take the opportunity to include in our graphics (see Fig.\ 3) the
recent results of \cite{BZ04} ({\ding{108}}) and further extend our
discussion of constraints on the Gegenbauer coefficients. \\
\begin{figure}[bt]
\label{fig:cleo-analysis}
 $$\includegraphics[width=0.49\textwidth]{%
 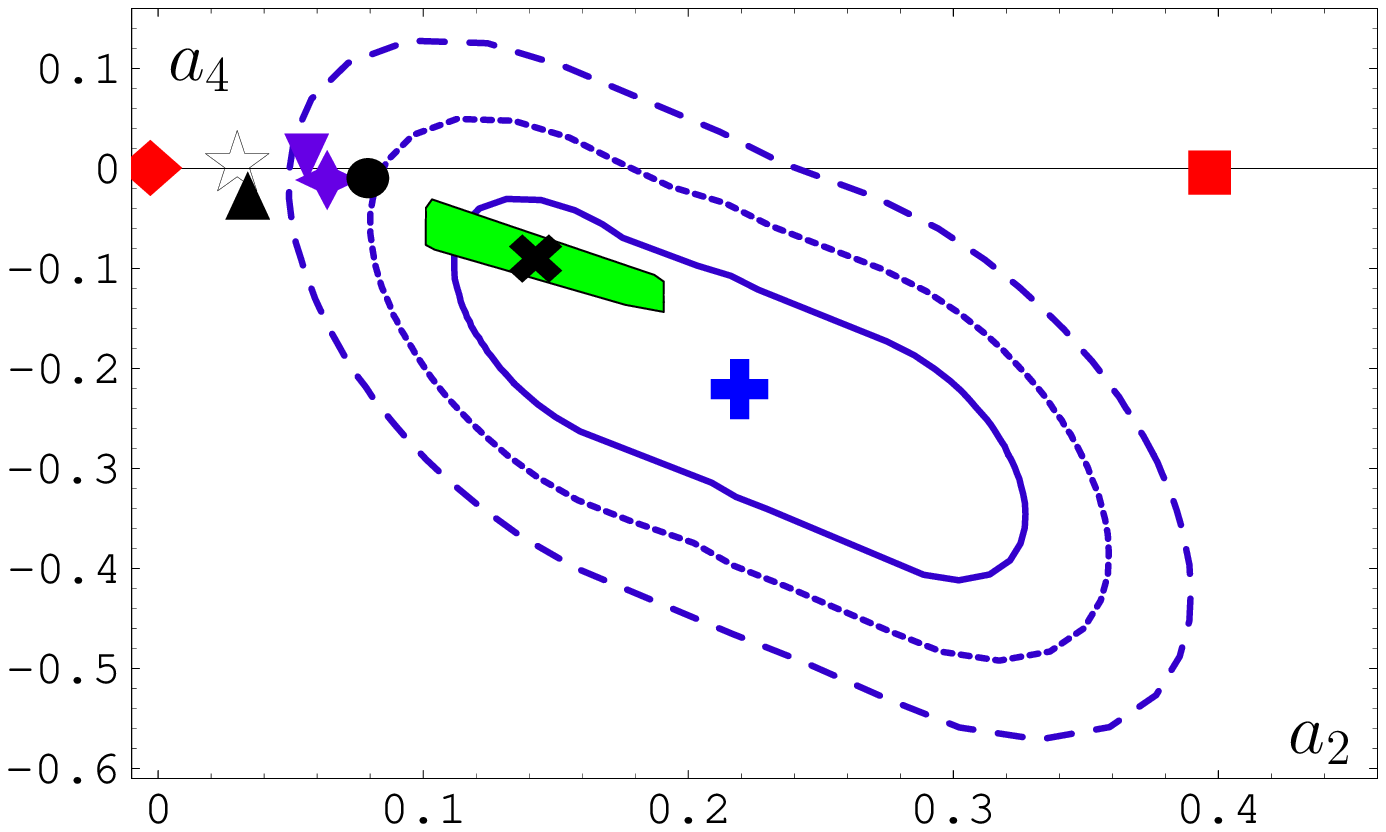}~~~
   \includegraphics[width=0.49\textwidth]{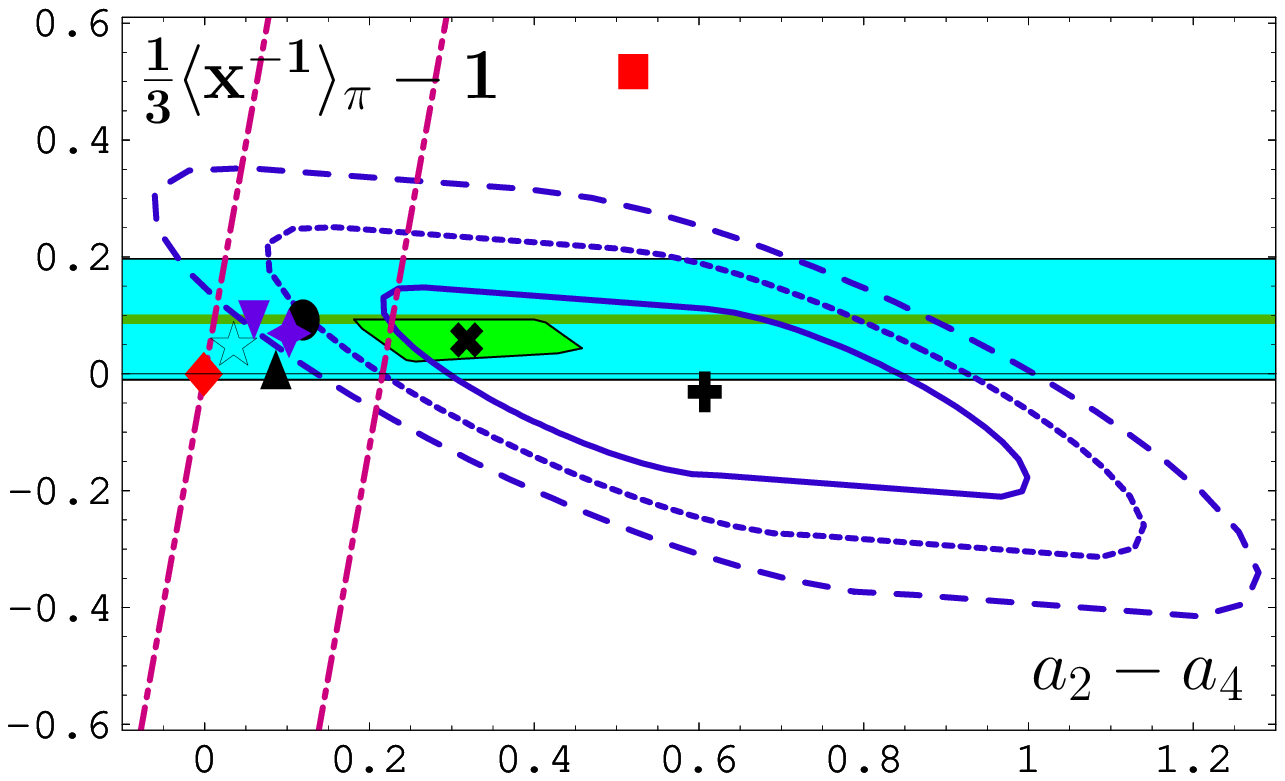}$$
\vspace*{-2mm}
 \caption{\footnotesize \textbf{Left:} Analysis of the CLEO data on
  $F_{\pi\gamma^{*}\gamma}(Q^2)$ in the ($a_2$, $a_4$) plane in terms
  of error regions around the best-fit point, using for both panels the
  following designations:
  $1\sigma$ (solid line);
  $2\sigma$ (dashed line);
  $3\sigma$ (broken line).
  Various theoretical models are shown:
  {\ding{117}}---asymptotic DA,
  {\ding{54}}---BMS model,
  {\footnotesize\ding{110}}---{CZ} DA,
  {\ding{58}}---best-fit point,
  {\ding{73} \protect{\cite{PPRWG99}}},
  {\ding{70} \protect{\cite{PR01}}},
  {\small\ding{115}} \cite{ADT00}---instanton models,
  {\ding{108}} \protect{\cite{BZ04}}---Ball-Zwicky (BZ) model
  \protect\cite{BZ04}, and
  {\footnotesize\ding{116}}---transverse lattice result
  \protect{\cite{Dal02}}.
  The slanted green rectangle represents the BMS ``bunch''
  of pion DAs dictated by the nonlocal QCD sum rules
  for the value $\lambda^2_q=0.4$~GeV$^{2}$.
  All constraints are evaluated at $\mu^2=5.76$~GeV$^2$
  after NLO ERBL evolution.
   \textbf{Right:}
   Results of the CLEO- data processing for
        $ \langle x^{-1} \rangle^\text{exp}_{\pi}/3-1$ at
        $\mu^2_0 \approx 1$ GeV$^{2}$ against
        theoretical predictions from QCD sum rules --
        hatched horizontal strip.
        The notations are the same as in the left panel.}
\end{figure}
Let us summarize the main findings of our analysis taking recourse
to figures 3. 
(i) The asymptotic DA and the CZ model are excluded at $3\sigma$ and
    $4\sigma$, respectively---as pointed out before by SY in
    \cite{SY99}. Several other proposed model DAs, extracted, for
    instance, from instanton-based approaches \cite{PPRWG99,PR01,ADT00},
    or from lattice calculations \cite{Dal02} are also disfavored---at
    least at the level of $2\sigma$, with the recently proposed model
    by Ball and Zwicky \cite{BZ04} lying exactly on the boundary of the
    $2\sigma$ error ``ellipse''. The important observation here is that
    only the BMS ``bunch'' lies entirely inside the $1\sigma$ error
    area of the CLEO data. A similar picture arises also for the
    inverse moment
    $\langle x^{-1} \rangle^{\rm exp}_{\pi}/3-1$ (right side of
    Fig.\ 3).
    The light solid line inside the hatched band indicates the mean
    value of the SR estimate,
    $\langle x^{-1} \rangle^\text{SR}_{\pi}/3-1=0.09$,
    and its error bars correspondingly.
    The strip bounded by two almost vertical dash-dotted lines
    corresponds to the rather old Braun--Filyanov \cite{BrFil89}
    constraints:
    $\varphi_{\pi}(1/2; \mu^2_0)=1.2 \pm 0.3$.
(ii) The extracted  parameters $a_{2}$ and $a_{4}$ were found to be
     rather sensitive to the strong radiative corrections and to the
     size of the twist-4 contribution. Nevertheless, even assuming a
     twist-4 uncertainty of the order of $30\%$, does not change these
     findings qualitatively. Still, both $\varphi_{\rm asy}$ and
     $\varphi_{\rm CZ}$ are outside the $3\sigma$ region with a slight
     improvement for the other models from $3\sigma$ to $2\sigma$.
(iii) The correlation length in the QCD vacuum was extracted directly
      from the CLEO data \cite{BMS02} and found to be
      $\Lambda\sim 0.31$~fm, i.e.,
      $\lambda_{\rm q}^{2}\lesssim 0.4$~GeV$^{2}$.
\begin{figure}[t]
 $$\includegraphics[width=0.49\textwidth]{%
 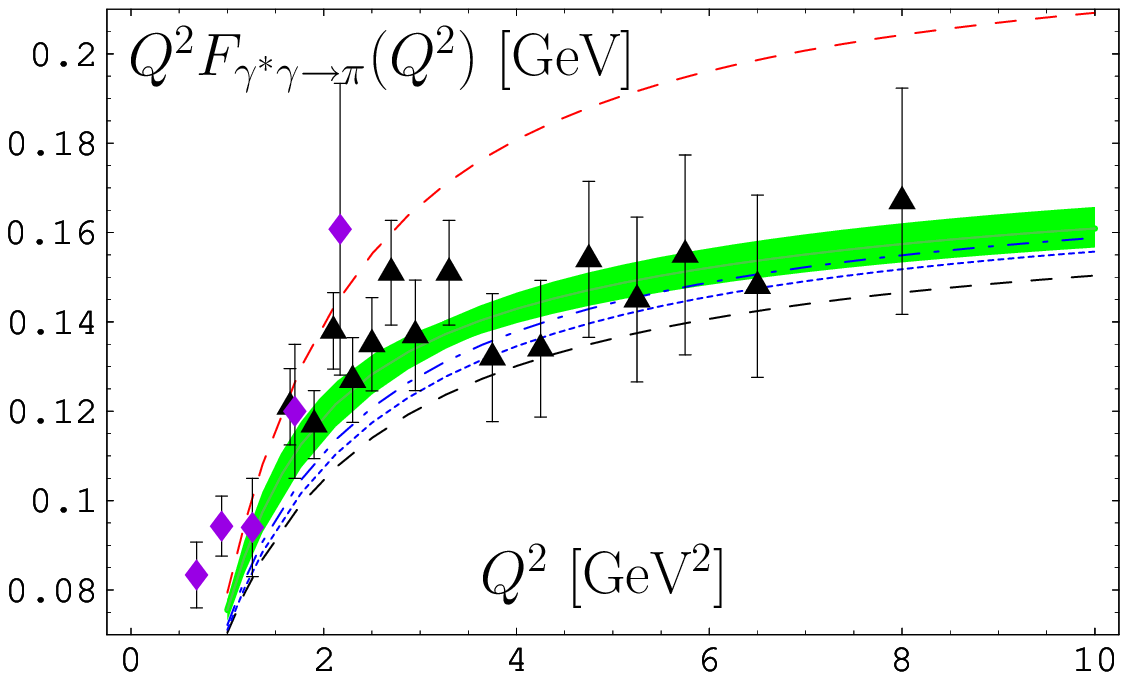}~~~
 \includegraphics[width=0.49\textwidth]{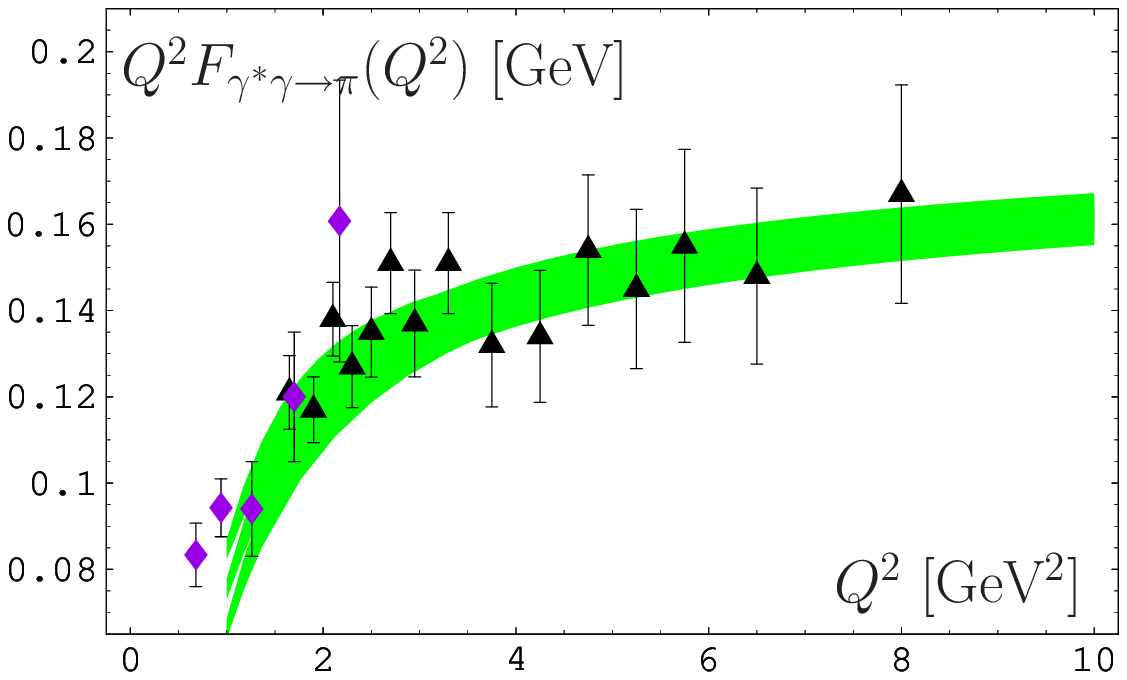}$$
 \vspace*{-10mm}
 \caption{\footnotesize
 \textbf{Left:} Light-cone sum-rule predictions for
   $Q^2F_{\gamma^*\gamma\to\pi}(Q^2)$
   in comparison with the CELLO (diamonds, \protect\cite{CELLO91})
   and the CLEO (triangles, \protect\cite{CLEO98}) experimental data,
   evaluated with the twist-4 parameter value
   $\delta_{\rm Tw-4}^2=0.19$~GeV$^2$~\protect\cite{BMS02,BMS03}.
   The predictions correspond to selected pion DAs; notably,
   $\varphi_{\rm CZ}$ (upper dashed line) \protect\cite{CZ84},
   BMS-``bunch'' (shaded strip) \protect\cite{BMS01},
   two instanton-based models, viz., \cite{PPRWG99} (dotted line)
   and \cite{PR01} (dash-dotted line),
   and $\varphi_{\rm as}$ (lower dashed line).
   \textbf{Right:} Similar predictions as in the left panel, but with
   the twist-4 parameter $\delta_{\rm Tw-4}^2$ varied in the
   range $\delta_{\rm Tw-4}^2=0.15-0.23$~GeV$^2$.
   }
\label{fig:pigammaFF}
\end{figure}
The prediction for the pion-photon transition form factor emerging
from this analysis is shown in Fig.\ \ref{fig:pigammaFF}. The
left-hand side shows the result for the twist-4 parameter
$\delta_{\rm Tw-4}^2=0.19$~GeV$^2$, while the right-hand side
illustrates the influence on the results of the variation of this
parameter in the range $\delta_{\rm Tw-4}^2=(0.15-0.23)$~GeV$^2$.
As one sees, in both cases, the prediction for the BMS ``bunch''
(shaded strip) is quite robust and in good agreement with both
sets of data---even at relatively low $Q^2$ values which exceed
the validity of the theoretical framework applied.

\section{Pion's electromagnetic form factor}
\begin{figure}[ht]
 $$\includegraphics[width=0.89\textwidth]{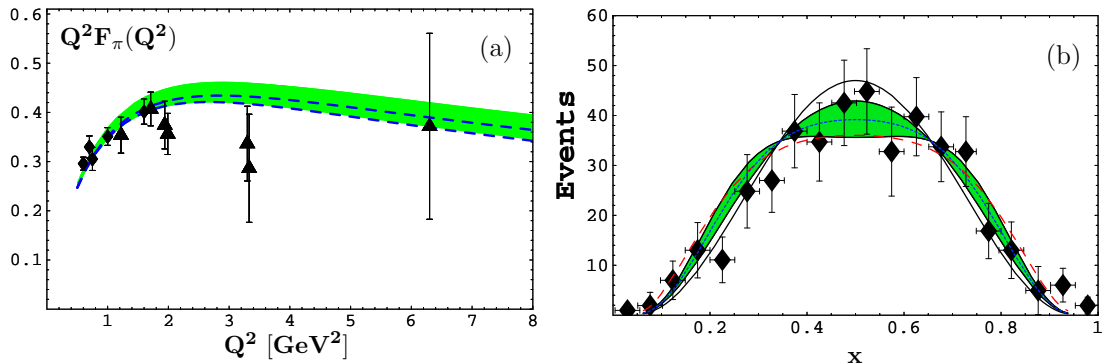}$$
  \caption{\footnotesize (a) Predictions for the scaled pion form
        factor calculated with the BMS\ bunch (green strip)
        encompassing nonperturbative uncertainties from nonlocal QCD
        sum rules \protect{\cite{BMS01}} and renormalization scheme and
        scale ambiguities at the level of the NLO accuracy.
        The dashed lines inside the strip restrict the area of
        predictions accessible to the asymptotic pion DA using the
        ``Maximally Analytic'' procedure \protect\cite{BPSS04}.
        The experimental data are taken from \protect{\cite{JLab00}}
        (diamonds) and \cite{FFPI73} (triangles).
        (b): Comparison with the E791 data \protect\cite{E79102}
        on diffractive di-jet production of the BMS ``bunch''
        (shaded strip), the asymptotic DA (solid line), and the CZ
        (dashed line) model, using the convolution approach of
        \cite{BISS02}.
\label{fig:pidatasum}}
\end{figure}

We have calculated in \cite{BPSS04} the electromagnetic pion form
factor
\begin{eqnarray}
  F_{\pi}(Q^{2};\mu_{\rm R}^{2})
  =  F_{\pi}^{\rm LD}(Q^{2})
  +  F_{\pi}^{\rm Fact-WI}(Q^2;\mu_{\rm R}^{2})\,,
\label{eq:Q2Pff}
\end{eqnarray}
where the soft part $F_{\pi}^\text{LD}(Q^{2})$ is modelled via
local duality and the factorized contribution
\begin{eqnarray}
 \label{eq:Fpi-Mod}
  F_{\pi}^{\rm Fact-WI}(Q^2;\mu_{\rm R}^{2})
  &=& \left(\frac{Q^2}{2s_0^{\rm 2-loop}+Q^2}\right)^2
       F_{\pi}^{\rm Fact}(Q^2;\mu_{\rm R}^{2})
\end{eqnarray}
with $s_0^{\rm 2-loop} \simeq 0.6$~GeV${}^2$
has been corrected via a power-behaved pre-factor in order to
respect the Ward identity at $Q^2=0$. In our analysis
$F_{\pi}^{\rm Fact}(Q^2;\mu_{\rm R}^{2})$ has been computed to NLO
\cite{MNP99,DR81}, using Analytic Perturbation Theory
\cite{Shi-an,SS-an} and trading the running coupling and its
powers for analytic expressions in a non-power series expansion,
i.e.,
\begin{eqnarray}
 \!\!\!\!\!\!\left[F_{\pi}^{\rm Fact}(Q^2; \mu_{\rm R}^{2})
             \right]_{\rm MaxAn}
 \ =\bar{\alpha}_{\rm s}^{(2)}(\mu_{\rm R}^{2})\,
 {\cal F}_{\pi}^{\rm LO}(Q^2)
   + \frac{1}{\pi}\,
      {\cal A}_{2}^{(2)}(\mu_{\rm R}^{2})\,
       {\cal F}_{\pi}^{\rm NLO}(Q^2;\mu_{\rm R}^{2})\,,
\label{eq:pffMaxAn}
\end{eqnarray}
with $\bar{\alpha}_{\rm s}^{(2)}$ and
${\cal A}_{2}^{(2)}(\mu_{\rm R}^{2})$
being the 2-loop analytic images of $\alpha_{\rm s}(Q^2)$
and $\left(\alpha_{\rm s}(Q^2)\right)^2$,
correspondingly (see \cite{BPSS04} and further references cited
therein), whereas
${\cal F}_{\pi}^{\rm LO}(Q^2)$ and ${\cal F}_{\pi}^{\rm
NLO}(Q^2;\mu_{\rm R}^{2})$ are the LO and NLO parts of the
factorized form factor, respectively. The phenomenological upshot
of this analysis is presented in Fig.\ \ref{fig:pidatasum}(a), where
we show $F_{\pi}(Q^{2})$ for the BMS ``bunch'' and using the
``Maximally Analytic'' procedure, which improves the previously
introduced \cite{SSK99} ``Naive Analytic'' one. This new procedure
replaces the running coupling and its powers by their analytic
versions, each with its own dispersive image \cite{Shi-an,SS-an}
and provides results in rather good agreement with the
experimental data \cite{FFPI73,JLab00}, given also the large
errors of the latter. One appreciates that the form-factor
predictions are only slightly larger than those resulting with the
asymptotic DA.

\section{Conclusions}
We have discussed pion DAs, derived from nonlocal QCD sum rules
that implement the idea of a nonzero vacuum correlation length,
and extracted for this latter quantity a value around $\sim
0.31$~fm directly from the CLEO data. The same set of experimental
data offers the possibility to determine the Gegenbauer
coefficients $a_2$ and $a_4$ within restricted error regions.
Using a data processing, based on light-cone sum rules, we have
shown that only the BMS-like pion DAs lie within the $1\sigma$
region, while all other known models are excluded at least at
$2\sigma$ or more. We have given theoretical predictions for the
electromagnetic form factor using NLO Analytic Perturbation Theory
and shown that the same BMS ``bunch'' of pion DAs provides results
very close to those obtained with the asymptotic DA. This
agreement proves that whether or not the pion DA is double-humped
or single-peaked is much less relevant compared to its endpoint
behavior. In the BMS case, the root cause for the excellent
agreement with the CLEO data is its strong endpoint suppression.
These findings---gathered in Table \ref{tab:piDAsvsData}---are
further backed up by the E791 data on diffractive di-jet
production, albeit this set of data alone cannot favor one DA over
the other. However, the fact that the middle region of $x$ where
the ``bunch'' has its largest uncertainties is within this data
range, provides independent credibility for this type of pion DAs
putting it into a wider general context. Finally, with the
approved upgrade of the CEBAF accelerator at JLab, planned to
provide precision data for the pion's electromagnetic form factor
up to $Q^2=6$~GeV$^2$, we expect that the confrontation between
theory and experiment will make a decisive step forwards.

\begin{table}[h]
\caption{Pros and cons of selected pion DAs (asymptotic-like; BMS\
\protect{\cite{BMS01}}; CZ \protect{\cite{CZ84}}) in comparison
with available experimental data.}
\begin{tabular}{cccc} \hline
$\pi$ DA models   & asymptotic-like
        & BMS \protect\cite{BMS01}\
           & CZ \protect\cite{CZ84}\
             \\ \hline \hline
$\pi-\gamma$ CLEO data \protect{\cite{CLEO98}}  & $3\sigma$~off
        & within~$1\sigma$
& $4\sigma$~off
              \\
JLab\ $F_\pi$ data \protect{\cite{JLab00,Blok02}}   & OK
        & OK
           & too large
             \\
Fermilab E791 \protect{\cite{E79102}} & $\chi^2=12.56$
        & $\chi^2=10.96$
           & $\chi^2=14.15$ \\ \hline
\end{tabular}
\label{tab:piDAsvsData}
\end{table}
\acknowledgments
It is a great pleasure to use this opportunity to wish Klaus Goeke
to preserve his vigor and activity for the years to come. This
work was supported in part by the Deutsche Forschungsgemeinschaft
(436 RUS 113/752/0-1), the Heisenberg--Landau Programme (grants
2002 and 2003), the Russian Foundation for Fundamental Research
(grants No.\ 03-02-16816 and 03-02-04022), and the INTAS-CALL 2000
N 587.

\end{document}